\documentclass[aps,prd,twocolumn,showpacs,nofootinbib]{revtex4}

\usepackage{amssymb} \usepackage{amsmath} \usepackage{graphicx}
\usepackage{epsfig,latexsym}
\RequirePackage{xspace} \allowdisplaybreaks

\begin{document}

\def\bef{\begin{figure}}
\def\eef{\end{figure}}

\newcommand{\nl}{\nonumber\\}

\newcommand{\ans}{ansatz }
\newcommand{\be}[1]{\begin{equation}\label{#1}}
\newcommand{\beq}{\begin{equation}}
\newcommand{\ee}{\end{equation}}
\newcommand{\beqn}[1]{\begin{eqnarray}\label{#1}}
\newcommand{\eeqn}{\end{eqnarray}}
\newcommand{\bd}{\begin{displaymath}}
\newcommand{\ed}{\end{displaymath}}
\newcommand{\mat}[4]{\left(\begin{array}{cc}{#1}&{#2}\\{#3}&{#4}
\end{array}\right)}
\newcommand{\matr}[9]{\left(\begin{array}{ccc}{#1}&{#2}&{#3}\\
{#4}&{#5}&{#6}\\{#7}&{#8}&{#9}\end{array}\right)}
\newcommand{\matrr}[6]{\left(\begin{array}{cc}{#1}&{#2}\\
{#3}&{#4}\\{#5}&{#6}\end{array}\right)}
\newcommand{\cvb}[3]{#1^{#2}_{#3}}
\def\lsim{\raise0.3ex\hbox{$\;<$\kern-0.75em\raise-1.1ex
e\hbox{$\sim\;$}}}
\def\gsim{\raise0.3ex\hbox{$\;>$\kern-0.75em\raise-1.1ex
\hbox{$\sim\;$}}}
\def\abs#1{\left| #1\right|}
\def\simlt{\mathrel{\lower2.5pt\vbox{\lineskip=0pt\baselineskip=0pt
           \hbox{$<$}\hbox{$\sim$}}}}
\def\simgt{\mathrel{\lower2.5pt\vbox{\lineskip=0pt\baselineskip=0pt
           \hbox{$>$}\hbox{$\sim$}}}}
\def\unity{{\hbox{1\kern-.8mm l}}}
\newcommand{\eps}{\varepsilon}
\def\ep{\epsilon}
\def\ga{\gamma}
\def\Ga{\Gamma}
\def\om{\omega}
\def\omp{{\omega^\prime}}
\def\Om{\Omega}
\def\la{\lambda}
\def\La{\Lambda}
\def\al{\alpha}
\newcommand{\ov}{\overline}
\renewcommand{\to}{\rightarrow}
\renewcommand{\vec}[1]{\mathbf{#1}}
\newcommand{\vect}[1]{\mbox{\boldmath$#1$}}
\def\tm{{\widetilde{m}}}
\def\mcirc{{\stackrel{o}{m}}}
\newcommand{\Dm}{\Delta m}
\newcommand{\dm}{\varepsilon}
\newcommand{\tanb}{\tan\beta}
\newcommand{\nbar}{\tilde{n}}
\newcommand\PM[1]{\begin{pmatrix}#1\end{pmatrix}}
\newcommand{\up}{\uparrow}
\newcommand{\down}{\downarrow}
\def\omE{\omega_{\rm Ter}}

%
%%%%%%%%%%     mauri    %%%%%%%%%%%%%%%%%%%%%%%%%%%%%%%%%

\newcommand{\Dsusy}{{susy \hspace{-9.4pt} \slash}\;}
\newcommand{\DCP}{{CP \hspace{-7.4pt} \slash}\;}
\newcommand{\mc}{\mathcal}
\newcommand{\gr}{\mathbf}
\renewcommand{\to}{\rightarrow}
\newcommand{\gtc}{\mathfrak}
\newcommand{\wh}{\widehat}
\newcommand{\br}{\langle}
\newcommand{\kt}{\rangle}

%%%%%%%%%%%%%%%%%%%%%%%%%%%%%%%%%%%%%%%%%%%%%%%%%%%%%%%%%%

\def\lsim{\mathrel{\mathop  {\hbox{\lower0.5ex\hbox{$\sim$}
\kern-0.8em\lower-0.7ex\hbox{$<$}}}}}
\def\gsim{\mathrel{\mathop  {\hbox{\lower0.5ex\hbox{$\sim$}
\kern-0.8em\lower-0.7ex\hbox{$>$}}}}}
%%%%%%%%%%%%%%%%%%%%%%%%%%%%%%%%%%

\def\nn{\\  \nonumber}
\def\de{\partial}
\def\brf{{\mathbf f}}
\def\bbf{\bar{\bf f}}
\def\bF{{\bf F}}
\def\bbF{\bar{\bf F}}
\def\bA{{\mathbf A}}
\def\bB{{\mathbf B}}
\def\bG{{\mathbf G}}
\def\bI{{\mathbf I}}
\def\bM{{\mathbf M}}
\def\bY{{\mathbf Y}}
\def\bX{{\mathbf X}}
\def\bS{{\mathbf S}}
\def\bb{{\mathbf b}}
\def\bh{{\mathbf h}}
\def\bg{{\mathbf g}}
\def\bla{{\mathbf \la}}
\def\bmu{\mathbf m }
\def\by{{\mathbf y}}
\def\bmu{\mbox{\boldmath $\mu$} }
\def\bsig{\mbox{\boldmath $\sigma$} }
\def\bunity{{\mathbf 1}}
\def\cA{{\cal A}}
\def\cB{{\cal B}}
\def\cC{{\cal C}}
\def\cD{{\cal D}}
\def\cF{{\cal F}}
\def\cG{{\cal G}}
\def\cH{{\cal H}}
\def\cI{{\cal I}}
\def\cL{{\cal L}}
\def\cN{{\cal N}}
\def\cM{{\cal M}}
\def\cO{{\cal O}}
\def\cR{{\cal R}}
\def\cS{{\cal S}}
\def\cT{{\cal T}}
\def\eV{{\rm eV}}

%
%%%%%%%%%%%%%%%%%%%%%%%%%%%%%%%%%%%%%

\title{Testing Noncommutative Spacetimes and Violations of the Pauli Exclusion Principle with underground experiments}

\author{Andrea Addazi$^1$}\email{andrea.addazi@infn.lngs.it}
\author{Pierluigi Belli$^{2,}$$^3$}\email{pierluigi.belli@roma2.infn.it}
\author{Rita Bernabei$^{2,}$$^3$}\email{rita.bernabei@roma2.infn.it}
\author{Antonino Marcian\`o$^1$}\email{marciano@fudan.edu.cn}

\affiliation{$^1$Center for Field Theory and Particle Physics \& Department of Physics, Fudan University, 200433 Shanghai, China}
\affiliation{$^2$INFN sezione Roma ``Tor Vergata", I-00133 Rome, Italy, EU}
\affiliation{$^3$Dipartimento di Fisica, Universit\`a di Roma ``Tor Vergata", I-00133 Rome, Italy, EU}

\begin{abstract}
\noindent
We propose to deploy limits that arise from different tests of the Pauli Exclusion Principle in order:
i) to provide theories of quantum gravity with an experimental guidance;
ii) to distinguish among the plethora of possible models the ones that are already ruled out by current data;
iii) to direct future attempts to be in accordance with experimental constraints. We firstly review experimental 
bounds on nuclear processes forbidden by the Pauli Exclusion Principle, which have been derived 
by several experimental collaborations making use of different detector materials. Distinct features of the 
experimental devices entail sensitivities on the constraints hitherto achieved that may differ one another by several orders of magnitude. 
We show that with choices of these limits, renown examples of flat noncommutative 
space-time instantiations of quantum gravity can be heavily constrained, and eventually ruled out. We devote particular attention 
to the analysis of the $\kappa$-Minkowski and $\theta$-Minkowski noncommutative spacetimes. These are deeply connected to 
some scenarios in string theory, loop quantum gravity and noncommutative geometry. We emphasize that 
the severe constraints on these quantum spacetimes, although cannot rule 
out theories of top-down quantum gravity to whom are connected in various 
way, provide a powerful limitations of those models that it will make sense to focus on in the future.

\end{abstract}

\maketitle

\section{Introduction}
\noindent 
The Pauli Exclusion Principle (PEP) is a direct implication of the Spin Statistics theorem (SST) stated by Pauli in Ref.~\cite{Pauli:1940zz}. PEP automatically arises from anti-commutation properties of fermionic 
creation and annihilation operators in the construction of the Fock space of the theory. In turn, the SST was proven by assuming Lorentz invariance. This certainly implies that the PEP is closely connected to the structure of the space-time itself. The PEP is indeed a successful fundamental principle not only when addressed from theoretical quantum field theory considerations, but also in high precision agreement with every atomics, nuclear and particle physics experimental data. In other words, if the PEP is violated, the violating channels must be parametrized by very tiny coupling constants in front of the PEP-violating operators. This possibility was suggested within an effective field theory approach in Refs.~\cite{Messiah:1900zz,Greenberg:1963kk,Gentile,Green:1952kp,Ignatiev:1987zd,Gavrin:1988nu,PEP1,PEP2,PEP3}.

The possibility of renormalizable PEP-violating operators might be seen as ``un-aesthetic'' and un-natural. However, the possibility of non-renormalizable effective operators induced by a PEP-violating new physics scale is still an open and natural possibility, which is predicted by many possible models of quantum gravity realizing an ultraviolet completion. A possible way to violate the PEP is, of course, to relax the main hypothesis on the basis of the Spin Statistics theorem. For example, as mentioned above, the theorem in its standard enunciation --- namely in terms of commutation relation for bosonic ladder operators, and anticommutation relation for fermionic ladder operators --- is no more valid if Lorentz invariance is relaxed. Lorentz symmetry is one of the basis of the Standard Model of particle physics: its explicit violation must allow any possible Lorentz Violating and CPT violating renormalizable operators. Even finetuned to very small couplings, the latter operators will introduce new UV divergent diagrams in the Standard Model sector, affecting the basic requirement of unitarity of the theory. This is why the Spin Statistics theorem, as a companion of Lorentz symmetry, is considered a {\it milestone} of the Standard Model. Notice furthermore, as pointed out in Ref.~\cite{Collins:2004bp}, that Lorentz violating effects --- for instance induced by the Planck scale in quantum gravity --- might manifest themselves in the propagation of low-energy particles with a sizable magnitude that in some cases is already ruled out by experimental data\footnote{It was shown in Ref.~\cite{Collins:2004bp} that only a strong and unnatural finetuning of the bare parameters of the theory may prevent from Lorentz violations at the percent level. Nonetheless, this analysis anyway does not take into account the possibility of a deformation of the Lorentz symmetries.}. 

Nonetheless, the eventuality that the Lorentz Symmetry is dynamically or spontaneously broken at a very high energy scale $\Lambda_{UV}$ is still open, and this must turn into the generation of non-renormalizable operators suppressed as inverse powers of $\Lambda_{UV}$. For example, many quantum gravity theories predict a {\it noncommutative} space-time geometry close to the Planck length scale. The idea that the space-time can be noncommutative was accredited to {\it W. Heisenberg} in Ref.~\cite{Heisenberg} and elaborated later on in Refs.~\cite{Snyder,Yang}. After few decades it was realized that noncommutativity of space-time can be rediscovered within the context of both\footnote{The fact that it appears in both theories may not be just a coincidence, in light of a new ${\bf H}$-duality conjecture that has been recently formulated in Ref.~\cite{Addazi:2017qwt}.} string theory \cite{Frohlich:1993es,Chamseddine1,Frohlich:1995mr,Connes:1997cr,Seiberg:1999vs} and loop quantum gravity \cite{AmelinoCamelia:2003xp,Freidel:2005me,Cianfrani:2016ogm,Amelino-Camelia:2016gfx, NClimLQG,BRAM, Brahma:2017yza}. Besides these two frameworks, many other studies have shed light on the emergence of deformed symmetries as a feature of effective theories that can be derived from (more fundamental?) non-perturbative models of quantum geometry --- see {\it e.g.} Ref.~\cite{MarMod}. 

Several studies were hitherto devoted also to the analysis of the physical meaning of deformed symmetries in spacetime, as {\it e.g.} in Refs.~\cite{Agostini:2006nc,Arzano:2007gr,Arzano:2007ef,AmelinoCamelia:2007uy,AmelinoCamelia:2007vj,AmelinoCamelia:2007wk,AmelinoCamelia:2007rn,Marciano:2008tva,Marciano:2010jm}. Some possible phenomenological consequences were also investigated in Refs~\cite{AmelinoCamelia:2007zzb,Marciano:2010gq,AmelinoCamelia:2010zf,AmelinoCamelia:2012it}, at least for those cases of noncommutativity that are the most manageable, namely those specific classes of models deforming the Lorentz symmetry to a noncocommutative space-time deformed symmetry group called $\kappa$-Poincar\'e and $\theta$-Poincar\'e. Among these latter, there exists a specific class of models enjoying $\theta$-Poincar\'e symmetries, which can preserve the unitarity of S-matrix in the Standard Model sector. This comes with a restriction \cite{AlvarezGaume:2003mb} on the components of the space-time noncommutative matrix $\theta^{\mu\nu}$. Under these assumptions, noncommutative quantum field theories with the Groenewold-Moyal product will not lead to catastrophic violation of the probability conservation principle \cite{Majid:1996kd,Oeckl:2000eg,Chaichian:2004za,Aschieri:2005yw,Balachandran:2004rq}. 

For these $\theta$-deformed theories, and for the class of deformations that enjoy $\kappa$-Poincar\'e symmetries \cite{Arzano:2016egk}, it is possible to show that deformed versions of the CPT theorem exists, or anyway that a deformed notion of discrete C, P and T symmetries can be recovered. This entails the introduction of a deformed SST, which encodes deformed commutation and anticommutation quantization rules, and thus deviation from the standard CPT theorem, which is nevertheless predicted to be small \cite{Balachandran:2005eb}. These deviations consequently lead to a violation of the standard PEP. Furthermore, it was enlightened in a series of work that CPT violation does not necessarily lead to violations of Lorentz invariance \cite{Chaichian:2011fc}, and vice versa \cite{Chaichian:2002vw,Chaichian:2011fc}, dismantling the bases of a well celebrated no-go theorem that was instead predicting this link, based on standard relativistic quantum field theory. These results call for an investigations of PEP directly at the level of the Fock space of the theory, where the breakdown or deformation of the theory induces deviations from ordinary statistics. In such a large panorama of possibilities, an effective parametrization approach is highly motivated, in order to experimentally distinguish among different scenarios that are theoretically allowed. We will account for deviations from commutation/anti-commutation relations of the creation and annihilation operators, which act on the vacuum in the Fock space of the theories, and then show how the cases of the theories enjoying $\kappa$-Poincar\'e symmetries $\theta$-Poincar\'e lie in a specific class of parametrization that allows a phenomenological falsification of (standard) PEP violations.

\section{Parametrization} 
\noindent
Operatively, deviations from the PEP in the commutation/anti-commutation relations can be parametrized --- see {\it e.g.} Refs.~\cite{PEP1,PEP2,PEP3} --- by introducing a deviation function $q(E)$, {\it i.e.}
\begin{eqnarray}\label{aadagger}
&&a_{i}a_{j}^{\dagger}-q(E)a_{j}^{\dagger}a_{i}=\delta_{ij}\, , \\
&&q(E)=-1+\beta^{2}(E),\quad {\rm and \ finally} \quad \delta^{2}(E)=\beta^{2}(E)/2\, . \nonumber
\end{eqnarray}
Among the possible parameterizations of the function $\delta^{2}(E)$, we propose a class corresponding to models that, depending on the order in the inverse powers of the energy scale of Lorentz violation, are classified at the $k$-th order as 
\begin{eqnarray}
&&M_{k}:\,\,\,\delta^{2}(E)=c_{k}\frac{E^{k}}{\Lambda^{k}}+O(E^{k+1})\,.\nonumber
\end{eqnarray}
The phenomenological method we deploy here naturally takes into account, through an analytic expansion driven by dimensional analysis, the corrections to the standard statistics that may arise, in the infrared limit, from UV-complete quantum field theories. This parametrization can capture every possible first term of the power series expansions in $E/\Lambda$, for every possible deformation function $q(E)$ in Eq.~(\ref{aadagger}). In other words, constraints on $\delta(E)$ can be translated into constraints on the new physics scale within the framework of the $M_{n}$ parametrization.

\section{Limits on PEP violating processes by underground experiments} 

\noindent
In order to investigate the aforementioned models we start referring to results obtained by the
underground experiments. Fig. \ref{fig:limits} shows the most stringent limits on the relative strength ($\delta^2$) for the searched non-paulian transitions.

\begin{figure*}[!ht]
\centerline{ \includegraphics [width=2\columnwidth]{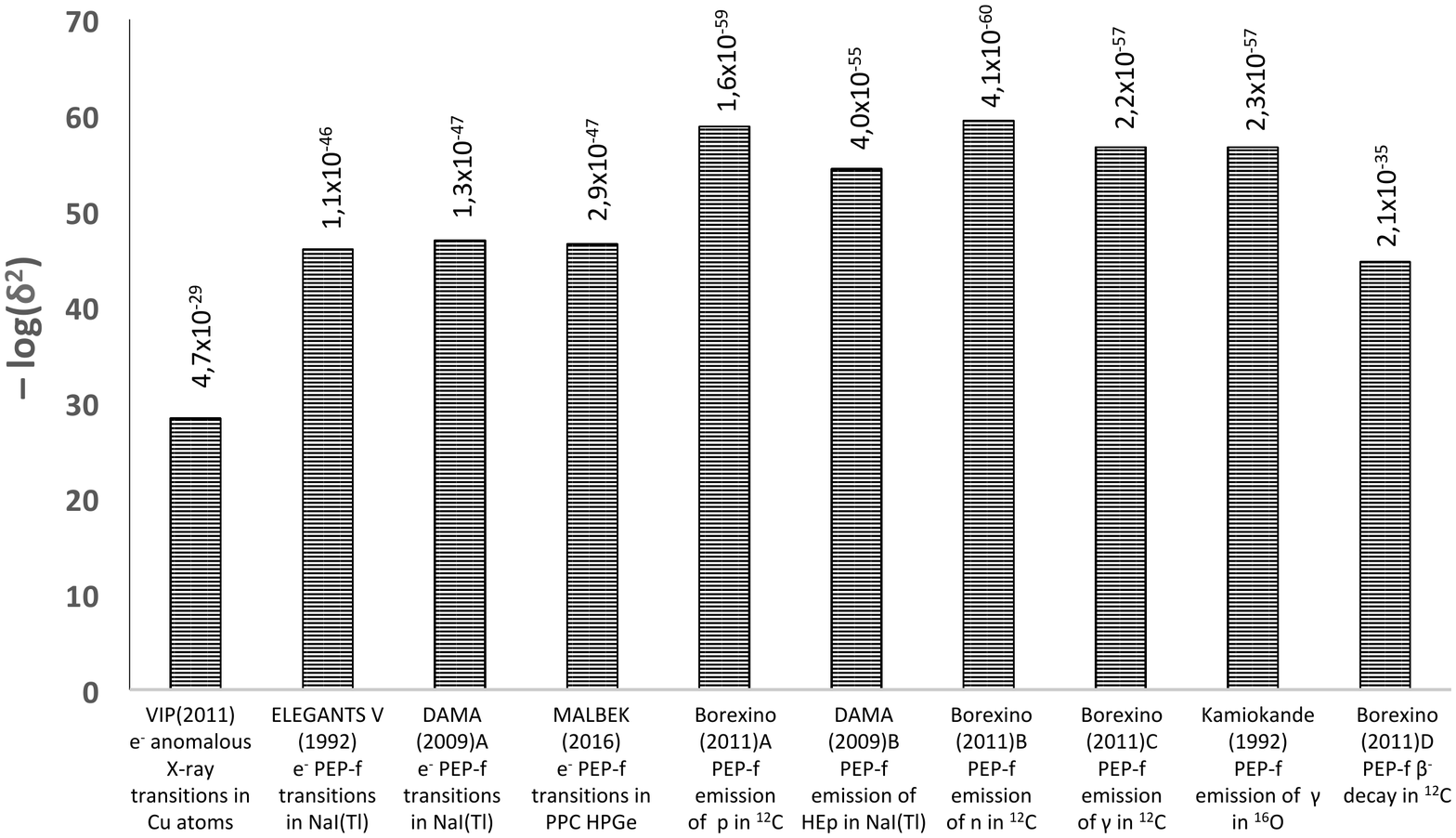}}
\vspace*{-20mm}
\caption{Limits at 90\% C.L. on various PEP violation channels in logarithmic scale, displaying the $-\log \delta^2$ for different experimental collaborations: VIP(2011) \cite{Pichler:2016xqc}; ELEGANTS V (1992) \cite{Ejiri:1992}; DAMA(2009)A \cite{Bernabei:2009zzb}; MALBEK(2016) \cite{Abgrall:2016wtk};
Borexino(2011)A \cite{Bellini:2010}; DAMA(2009)B \cite{Bernabei:2009zzb}; Borexino(2011)B \cite{Bellini:2010}; Borexino(2011)C \cite{Bellini:2010}; Kamiokande(1992) \cite{Suzuki:1993zp}; Borexino(2011)D \cite{Bellini:2010}. }
\label{fig:limits}
\end{figure*}

Several methods of experimental investigations for testing PEP have been used so far. The VIP experiment \cite{Pichler:2016xqc}
uses a method of searching for PEP forbidden atomic transitions in copper; the limits on the probability that PEP is violated by electrons
are reported in Fig. \ref{fig:limits}.
The experimental method consists in the injection of ``fresh" electrons into a copper strip, by means of a circulating current, and in the search for the X-rays following the possible PEP forbidden radiative transitions that occur if one of these electrons is captured by a copper atom and cascades down to the already-filled 1S state. In particular, the experiment is searching for the $K_\alpha$ (2P $\rightarrow$ 1S) transition. The energy of this PEP forbidden transition (7.729 keV) would differ from the normal $K_\alpha$ transition energy (8.040 keV) by a $\Delta$ term (about 300 eV) due to the presence
of the other electrons in the already-filled shell. This energy shift can be
detected by the high resolution CCD devices.

PEP forbidden radiative atomic transitions are also searched for in Iodine atoms deploying NaI(Tl) detectors, as done in DAMA/LIBRA (DAMA(2009)A in Fig. \ref{fig:limits}) \cite{Bernabei:2009zzb}
and ELEGANTS V \cite{Ejiri:1992} experiments, and in Germanium atoms in PPC HPGe detectors of the MALBEK experiment \cite{Abgrall:2016wtk} (see Fig. \ref{fig:limits}).
In such cases, when a PEP-violating electronic transition occurs, X-rays and Auger electrons are emitted
by the transition itself and by the following rearrangements of the atomic shell.
The detection efficiency of such radiation in the NaI(Tl) detectors of DAMA/LIBRA is $\simeq 1$ at the low energy of the process.
Thus, all the ionization energy for the considered shell is detected, but it is actually shifted by a $\Delta$ term due to the presence
of the other electrons in the already filled shells.
Generally, in this class of experiments the K-shell is considered, as it provides the largest available energy in the subsequent X-rays /Auger-electrons radiation emission.
However, stringent limits (not reported in Fig. \ref{fig:limits}) were also obtained by DAMA/NaI looking for transitions to L-shell in Iodine atoms \cite{bernabei2}, providing 4-5 keV radiation emission, thanks to
the low energy thresholds of such NaI(Tl) detectors.

The most stringent constraint on this class of PEP violations in atomic transitions comes from the DAMA/LIBRA experiment, a 250 kg array of highly radiopure NaI(Tl) detectors hosted in the Gran Sasso National Laboratory. DAMA/LIBRA searched for PEP violating K-shell transitions in Iodine using the data corresponding to 0.53 ton$\times$yr; a lower limit on the transition lifetime of $4.7 \times 10^{30}$ s has been set, giving $\delta^2 < 1.28 \times 10^{-47}$ at 90\% C.L. \cite{Bernabei:2009zzb}. This value is reported in Fig. \ref{fig:limits}.

A similar experiment, MALBEK, has been using a high-purity germanium (HPGe) detector with an energy threshold suitable for observing the transition from L- to K- shells in germanium. In this case, the energy of the transition has been calculated to be 9.5 keV \cite{Abgrall:2016wtk}, once shifted down by the $\Delta$ term. The obtained limit on $\delta^2$ is also reported in Fig. \ref{fig:limits}.

A different approach for studying PEP violating processes has been exploited by DAMA/LIBRA collaboration (DAMA(2009)B in Fig. \ref{fig:limits}) \cite{Bernabei:2009zzb}. Specifically, PEP violating transitions in nuclear shells of $^{23}$Na and $^{127}$I are investigated by studying possible protons emitted with E$_p \ge$ 10 MeV. In such a case, events with only one detector fires, that is each detector has all the others as veto, are considered to search for high energy protons. The rate of emission of high energy protons (E$_p \ge$ 10 MeV) due to PEP violating transitions in $^{23}$Na and $^{127}$I was constrained to be $\lsim 1.63 \times 10^{-33}$ s$^{-1}$ (90\% C.L.) \cite{Bernabei:2009zzb}. This corresponds to a limit on the relative strength of the searched PEP violating transitions: $\delta^2 \lsim 4 \times 10^{-55}$ at 90\% C.L. (see Fig. \ref{fig:limits}).

Moreover, the Pauli exclusion principle has also been tested with the Borexino detector \cite{Bellini:2010}, considering the nucleons in the $^{12}$C nuclei. The research profits of the extremely low background and of the large mass (278 tons) of the
Borexino detector.
The exploited method is to look for $\gamma$, $\beta^\pm$, neutrons, and protons, emitted in a PEP violating transition of nucleons from the 1P$_{3/2}$ shell
to the filled 1S$_{1/2}$ shell and the following limits on the lifetimes for the different PEP violating transitions were set \cite{Bellini:2010}
(all the limits are at 90\% C.L.):
$\tau$($^{12}$C $\rightarrow ^{12}\widetilde{C} + \gamma$) $\ge 5.0 \times 10^{31}$ yr,
$\tau$($^{12}$C $\rightarrow ^{12}\widetilde{N} + $e$^{-} + \bar{\nu}_e$) $\ge 3.1 \times 10^{30}$ yr,
$\tau$($^{12}$C $\rightarrow ^{12}\widetilde{B} + $e$^{+}  + \nu_e$)       $\ge 2.1 \times 10^{30}$ yr
$\tau$($^{12}$C $\rightarrow ^{11}\widetilde{B} + $p)      $\ge 8.9 \times 10^{29}$ yr, and
$\tau$($^{12}$C $\rightarrow ^{11}\widetilde{C} + $n)      $\ge 3.4 \times 10^{30}$ yr.

These limits correspond to constraints on the relative strengths for the searched PEP violating electromagnetic, strong and weak transitions:
$\delta^2_\gamma \le 2.2 \times 10^{-57}$,
$\delta^2_N      \le 4.1 \times 10^{-60}$, and
$\delta^2_\beta  \le 2.1 \times 10^{-35}$ (see Fig. \ref{fig:limits}) \cite{Bellini:2010}.

Finally, we report here the results obtained by the large underground water Cherenkov detector, Kamiokande \cite{Suzuki:1993zp}, where anomalous
emission of $\gamma$ rays in the energy range $19 - 50$ MeV has been searched for. No statistically significant excess was found
above the background; this allows to set a limit on
the lifetime of PEP violating transitions to $9.0 \times 10^{30} \times$ Br($\gamma$) yr per oxygen nucleus, where Br($\gamma$) is the branching ratio of the $^{16}$O decay in the $\gamma$ channel.
In the case the PEP violating transitions is due to the p-shell nucleons, then the limit is $1.0 \times 10^{32} \times$ Br($\gamma$) yr.
Thus, the limit at 90\% C.L. of the relative strength for forbidden transitions to normal ones is
$\delta^2 < 2.3 \times 10^{-57}$  \cite{Suzuki:1993zp}, it is also shown in Fig. \ref{fig:limits}.

\section{Implications to Planck-scale deformed symmetries} 
\noindent
We can start considering a generic model, with the assignments $M_{k}$, for $k\in \mathbb{N}$. On these latter, using the DAMA/LIBRA results as example, the 
following constraints can be derived:
\begin{eqnarray} \label{bound}
\delta^{2}\leq 4 \times 10^{-55} \longleftrightarrow \delta^{2}(E)=c_{k}\frac{E^{k}}{\Lambda^{k}}\leq 4\times 10^{-55}\,.
\end{eqnarray}
We are interested in those cases that are mostly motivated by quantum gravity scenarios. This corresponds to select $\Lambda = M_{Pl} \simeq 1.22\times 10^{19}\, {\rm GeV}$. 
A straightforward estimate of $k$ can be then achieved, which has already dramatic consequences for several models of quantum gravity. Since nuclear decays processes taking place in the detector have an energy whose order of magnitude is few times $10^{-3}$ GeV, we may consider $E = 10\,\pm1 \,{\rm MeV}$. For a set of heuristic choices of $c_k=\{1,4, 10\}$, this implies immediately that at 90\% C.L. only $k > k^*$ power suppressions are still experimentally allowed, with respectively\footnote{The propagation of the error only affects the last digit of $k^*$, and is effectively independent on these heuristic choices of $c_k$, which capture the range of values in the literature --- see {\it e.g.} Refs.~\cite{Arzano:2007ef,Freidel:2007hk,Balachandran:2004rq,Balachandran:2005eb,Daszkiewicz:2007ru,Arzano:2008yc,AmelinoCamelia:2007zzb}. On the other hand, the theoretical ambiguity on $c_k$ may affect the estimate of $k^*$ up to 2\% of its value.} $k^*=\{2.58, 2.61, 2.63 \}\pm 0.01\,$. The exclusion limits on the $k$--$\Lambda$ plane are displayed in Fig.~\ref{fig:pep}, in which we use the accurate values for $E$ that pertain the different experiments analyzed, and set the coefficients\footnote{As remarked in the previous footnote, this choice is motivated by the fact that in the literature about noncommutative spacetimes the $c_k$ coefficients are order $1$.} $c_k=1$. The most stringent constraints on the $k$--$\Lambda$ parameters' plane, obtained by the above-mentioned experimental limits on the relative strength for non-paulian transitions, are provided by the Borexino \cite{Bellini:2010}, Kamiokande \cite{Suzuki:1993zp} and DAMA \cite{Bernabei:2009zzb} collaborations.

A different scenario arises while working at a scale of energy  $E \simeq 10$ keV, which is induced by transitions from k-electronic shells. This  provides the upper bound $\delta^2 \simeq 10^{-47} \!-\! 10^{-48}$, which is less stringent than the former one. Nonetheless, it still entails rejection of PEP violating terms that are suppressed at the second order in $\Lambda$, and at the same time are regulated by coefficients $c_n$ of order one.

Below we focus on PEP violations that arise in the aforementioned models of noncommutative spacetime, with particular focus on models endowed with $\kappa$-Poincar\'e deformed symmetries, and $\theta$-Poincar\'e deformed symmetries. Notice that the latter can be recast in the language of the noncommutative geometries {\it \`a la} Connes 
\cite{Gayral:2003dm}, while for the former the equivalence has been proven hitherto under 
certain restrictions \cite{Iochum:2010hi,Matassa:2012su}. Besides the links to 
noncommutative geometry, noncommutative spacetimes have also been derived in several 
frameworks of quantum gravity, most notably in string theory and loop quantum gravity. In 
the former scenario the Groenewold-Moyal noncommutativity is induced by the expectation value of background $B$ fields \cite{Seiberg:1999vs}, while in the latter several instantiation of $\kappa$-deformation have been so far derived, with mesoscale deviations from Lorentz invariance --- see {\it e.g.} \cite{Freidel:2005me,Cianfrani:2016ogm,Amelino-Camelia:2016gfx,BRAM}. Although these derivations are not yet decisively answering the question about the low energy limit of string theory and loop quantum gravity, the constraints that we are providing here will have the undoubtable advantage to furnish a guidance for the development of theoretical models of quantum gravity. 

\begin{figure}[ht]
\centerline{ \includegraphics [width=0.98\columnwidth]{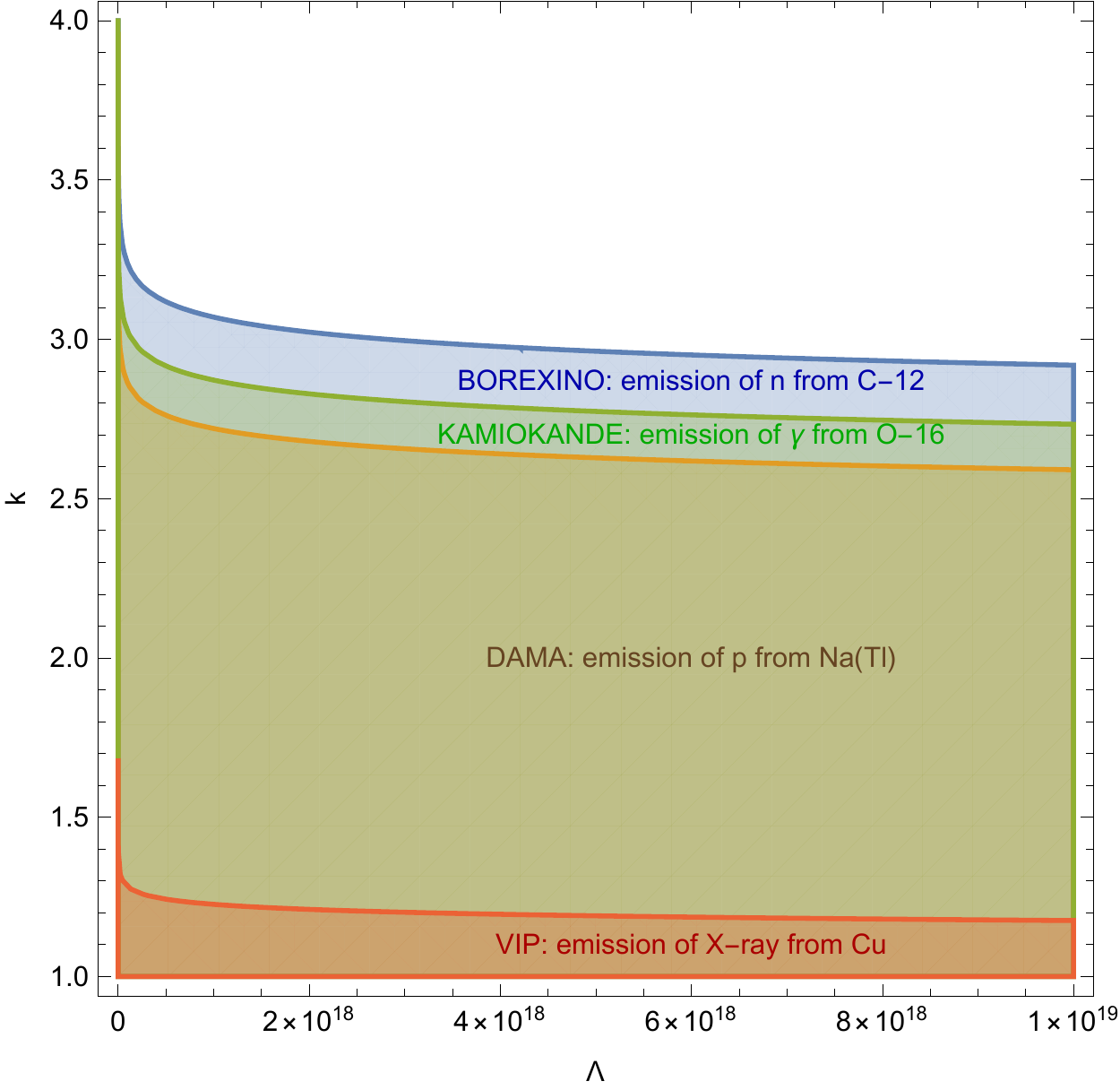}}
%\vspace*{-1ex}
\caption{Exclusion limits at 90\% C.L. on the $k$--$\Lambda$ parameters' plane. Most stringent constraints are provided by the Borexino \cite{Bellini:2010}, Kamiokande \cite{Suzuki:1993zp} and DAMA \cite{Bernabei:2009zzb} collaborations.}
\label{fig:pep}  
\end{figure}

\subsection{The case of the $\kappa$-Poincar\'e group }
\noindent 
%{\it The case of the $\kappa$-Poincar\`e group.} 
Small departures from locality --- an essential requirement for micro-causality in standard quantum field theory --- may be kinematically or dynamically generated in some quantum gravity scenarios, and have been shown in \cite{Arzano:2007nx} to be connected to the emergence of deformed $\kappa$-Poincar\'e symmetries. %Furthermore, several patterns --- see {\it e.g.} Ref.~\cite{BRAM} --- were enlightened by mean of which space-time symmetries $\kappa$-deformations may arise.  
This is a non-trivial Hopf-algebra of symmetries dual to the $\kappa$-Minkowski non-commutative spacetime. The latter is characterized by commutation relations among the spacetime point coordinates of the type
$$[ x_i,x_j]=0 \qquad  {\rm and} \qquad [ x_i,x_0]=\frac{\imath}{\kappa} x_i\,,$$ 
where $\kappa$ denotes a scale of energy assumed to be order of the Planck scale in effective quantum gravity frameworks. There exists at least a basis of the Hopf algebra in which the Lorentz sector is standard, but the action of the Lorentz generators on the translation subgroup is $\kappa$-deformed. For instance, in the {\it bicrossproduct basis} there is one $\kappa$-deformed  commutator, namely  
$$ [P_l, N_j]=  - \imath \delta_{lj} \left( \frac{\kappa}{2} \left( 1- e^{-\frac{2 P_0}{\kappa}}\right) \frac{1}{2\kappa} \vec{P}^2 \right) +\frac{\imath}{\kappa}P_l \, P_j\,,$$
where $P_\mu$ denote spacetime translation generators and $N_l$ boost generators. Even in this basis, in which translation generators remain commutative, the co-product map $\Delta$ acquires a $\kappa$-deformation. This is a remarkable deviation from standard properties. When $P_\mu$ are represented as derivatives acting on ``coordinates-ordered'' exponentials \cite{Arzano:2016egk}, it is trivial to recognize that $\Delta$ generalizes the Leibnitz rule of derivatives' action. The $\kappa$-deformed Leibnitz rule can then be inferred by the only $\kappa$-deformed co-product, {\it i.e.} $$\Delta(P_j)=P_j\otimes 1\!\!1 + e^{-\frac{P_0}{\kappa}} \otimes P_j\,.$$
The fate of discrete symmetries in the $\kappa$-Poincar\'e setting was addressed in Ref.~\cite{Arzano:2016egk}, while a detailed analysis of the fate of the CPT theorem for $\kappa$-Poincar\'e symmetries and of its consequences is still missing. Nonetheless, a phenomenological analysis of deviations from the standard case is still possible. Moving from the parametrization in Eq.~\eqref{aadagger}, by straightforward dimensional arguments we can express 
\begin{equation}
\delta^2 (E)=c_1 \kappa E\,,
\end{equation}
where it is assumed that $\kappa\simeq M_{Pl}^{-1}$. This implies automatically the rejection of every model available in the literature that predicts a $c_1$ non-vanishing and of order one. 

Following a constructive procedure, we can show that most part of the models hitherto addressed in the literature  --- see {\it e.g.} Refs.~\cite{Arzano:2007ef,Freidel:2007hk,Daszkiewicz:2007ru,Arzano:2008yc,AmelinoCamelia:2007zzb} --- either reproduce the case $c_1$ order one, or they fall in the class of a vanishing $c_1$, for which they cannot be falsified at the level of PEP violations. For instance, in Refs.~\cite{Arzano:2007ef,Arzano:2008yc,AmelinoCamelia:2007zzb} $c_1=1$, and consequently the models are ruled out. While in Refs.~\cite{Freidel:2007hk,Daszkiewicz:2007ru}, where  $c_k=0$, for $k\in \mathbb{N}$, the commutation relations are unmodified. This scenario can be then falsified up to the second order in the ratio $E/M_{Pl}$, but is not distinguishable from the standard unmodified case. 

\subsection{The case of the Groenewold-Moyal plane }
\noindent 
%{\it The case of the Groenewold-Moyal plane.} 
The algebra $\mathcal{A}(R^{d})$ of commutative functions on a smooth d-dimensional space-time manifold can be mapped into that one of noncommutative functions on the Groenewold-Moyal plane $\mathcal{A}_{\theta}(R^{d})$, if the star-product is considered 
%\begin{equation}
%\label{product}
$(\alpha * \beta)(x)=\left( \alpha\, e^{ \frac{\imath}{2} \  ^{\leftarrow}\partial_{\mu}\theta^{\mu\nu}\partial_{\nu}^{\rightarrow}} \beta \right)(x)\,,$ 
%\end{equation}
where $\theta^{\mu\nu}=-\theta^{\nu\mu}$ and $x=(x^{0},...\,,x^{d-1})$. Accordingly, the Groenewold-Moyal (GM) multiplication map $m_{\theta}$ reads
\begin{equation}
\label{mtheta}
m_{\theta}(\alpha \otimes \beta)=m_{0}\left( e^{ -\frac{\imath}{2} \theta^{\mu\nu} (-\imath\partial_{\mu})\otimes (-\imath\partial_{\nu}) } \cdot (\alpha \otimes \beta ) \,\right)\,,
\end{equation}
where $m_{0}(\gamma\otimes \delta)(x) \equiv \gamma(x)\delta(x)$ stands for the standard point-wise multiplication rule. 

Introducing the invertible element of the $\mathcal{R}$-matrix 
$$\mathcal{F}_{\theta}=e^{-\frac{\imath}{2}(-\imath\partial_{\mu})\theta^{\mu\nu}\otimes (-\imath\partial_{\nu})}\,,$$
the GM multiplication rule can be recast as 
%\begin{equation}
%\label{mFtheta}
$m_{\theta}(\alpha \otimes \beta)=m_{0}(\mathcal{F}_{\theta}\alpha \otimes \beta)\, .$ 
%\end{equation}
%
The invertible element of the $\mathcal{R}$-matrix enters in a natural way the twisted deformation of the Fock space of scalar field theory, with spin zero, and thus the commutation relations of the ladder operators, {\it i.e.}
\begin{eqnarray}
\label{product2}
a(p)a^{\dagger}(q)=\tilde{\eta}'(p,q)\tilde{F}_{\theta}^{-2}(-q,p)a^{\dagger}(q)a(p) \nonumber \\
+2p_{0}\delta^{d}(p-q)\,,
\end{eqnarray}
where $\tilde{\eta}'$ approaches the constant $+1$  in the low energy limit --- this is formally equivalent to the commutative limit $\theta^{\mu\nu}\rightarrow 0$. Anti-commutation relations for free spinor fields are equal to the ones given in Eq.~(\ref{product2}), provided that $\tilde{\eta}'$ approaches $-1$ in the low energy limit. 

We may expand Eq.~\eqref{product2} at first order in $\theta^{\mu\nu}$, neglecting orders $O(\theta_{\mu\nu}\theta^{\mu\nu})$. This corresponds to a second order expansion in $\Lambda$, since $\theta^{\mu\nu}$ has dimension of length square. This immediately entails $c_1\equiv 0$, and allows to set $c_2=1$ provided that $\theta^{12}=\theta^{13}=\theta^{23}=1/(3 \Lambda^{2})$. Focusing on the data provided by DAMA (2009) B \cite{Bernabei:2009zzb}, and accounting for an isotropic distribution of the protons' momenta, we obtain that  
$$\delta^2_{\theta}= \left(\frac{E}{\Lambda}\right)^2\,,$$ 
the exponent of which can be confronted with the values of $k$ excluded in Fig.~2. Thus this model seems to be already excluded by present data. This is as transparent as surprising, since it was never pointed out in the wide literature devoted to non-commutative space-times. 

\subsection{Quantum gravity with lower energy scales}
\noindent 
%{\it Quantum gravity with lower energy scales.} 
We can resort to the experimental bound in \eqref{bound} in order to constrain departures from the standard spin-statistics theorem within those theoretical frameworks that predict a lower energy scale of quantum gravity. Several models fit this scenario, notably the proposals that took into account an eventual role of large scale extra dimensions in the resolution of the hierarchy problem --- see {\it e.g.} 
Refs~\cite{ArkaniHamed:1998nn,ArkaniHamed:1998rs,Dvali:2001gm}. It is then straightforward to check that any violation of PEP could arise up to the ninth order in the ratio $E/\Lambda$, within those proposals where the scale of quantum gravity is lowered down to the threshold hitherto achievable on terrestrial experiments, $\Lambda\simeq 10$ TeV. This rules out any reliable model of extra dimension that would break Lorentz invariance and would predict violations of PEP.

\section{Conclusions and outlooks} 
\noindent 
Although a direct connection between deformation of space-time symmetries and quantum gravity has not yet been decisively proved, nonetheless there are many results in the literature that provide a clear instantiations of space-time symmetry deformation or space-time symmetry breakdown regulated by the Planck scale. Making contact in particular with those models that predict a deformation of the energy-momentum dispersion relations for one-particle states, and thus entail a deformation of the Fock space states and of the SST, we developed a framework to falsify these scenarios accounting for possible PEP violations. 

We emphasize that the phenomenological analysis we developed here differentiates from previous phenomenological investigations accounting for the one-particle Hilbert space structure of quantum field theories on non-commutative space-times. Constraints on energy-momentum dispersion relations do apply only to certain classes on non-commutative space-times. For instance, quantum field theories endowed with $\kappa$-Poincar\'e symmetries, in which the algebra and the mass Casimir are deformed, provide an arena to test deformations of the energy-momentum dispersion relations. On the other hand, quantum field theories endowed with $\theta$-Poincar\'e symmetries can be falsified only by looking at deformations of the Fock space structure, including eventual violations of the Pauli exclusion principle.

The tightest constraints on in-vacuo dispersion relations that are sensitive to the Planck-scale, as discussed within the phenomenological models in Refs.~\cite{C1,C2,C3,C4,C5,C6}, are provided for photons by the observation of TeV flares originated from active galactic nuclei at redshift smaller than $1$ --- see e.g. Refs.~\cite{C6,C7,C8,C9,C10,C11}. Taking into account deformation's effects that are linear in the Planck length scale, the bounds can reach $1/10$ of the Planck scale. 
On the other hand, the best constraints on anomalous in-vacuo dispersion that are quadratic in the Planck length-scale may be obtained from the detection of neutrinos\footnote{Indeed for these neutrinos there is also the possibility to compare the arrival times with that one of low-energy photons.} emitted by gamma ray bursts, with energies between $10^{14}$ and $10^{19}$ TeV --- see e.g. Refs.~\cite{C6,C12,C13,C14,C15,C16,C17,C18}.  
This clearly shows the relevance of our analysis with respect to the constraints previously discussed in the literature. Our analysis indeed provides either a restriction of the dimensionful parameters entering the UV-complete theories to be tested or a rejection/acceptance of their theoretical predictions. For instance, for string theory we can only restrict the values of the parameters involved in the theoretical construction, while in the case of loop quantum gravity, the only dimensionful scale is the Planck-scale, and all the order-one dimensionless parameters are fixed by the theory. Thus, with our analysis we were able to provide for all these attempts a restriction of the universality classes that are allowed on the theoretical side, and rule-out values of the parameters that are either the most natural ones --- from a theoretical perspective --- to be considered, or the only ones that can be considered.

Dedicated measurements can be planned in forthcoming updates of DAMA/LIBRA and other experiments. This may provide the chance of constraining $M_n$ with $n\geq 3$ contributions, which are suppressed by the $n$-th power of the energy scale $\Lambda$. In particular, an increase of sensitivity in $\delta^2$ would trigger the possibility of constraining third order suppressed PEP violating terms. 
%Finally, we mention that new results might arise from future experiments, such as JUNO \cite{juno}, where the increase on the sensitive mass might turn in an unprecedented increase of sensitivity in $\delta^2$. In this latter case, the exploited process will be to search for PEP violating transitions of nucleons in $^{12}$C nucleus from the 1P$_{3/2}$ shell to the filled 1S$_{1/2}$ shell, as Borexino already did. 
For completeness we also mention the potentiality of a very interesting result on this topic from data collected by Super-Kamiokande.

\vspace{2cm}

\acknowledgments
\noindent 
We wish to acknowledge G.~Amelino-Camelia and M.~Arzano for insightful comments on the draft.


\vspace{2cm}
\begin{thebibliography}{99}

% MODEL INDEPENDENT PEPV



%\cite{Pauli:1940zz}
\bibitem{Pauli:1940zz}
  W.~Pauli,
  %``The Connection Between Spin and Statistics,''
  Phys.\ Rev.\  {\bf 58} (1940) 716.
  doi:10.1103/PhysRev.58.716
  %%CITATION = doi:10.1103/PhysRev.58.716;%%
  %249 citations counted in INSPIRE as of 10 Sep 2017

%\cite{Messiah:1900zz}
\bibitem{Messiah:1900zz}
  A.~M.~L.~Messiah and O.~W.~Greenberg,
  %``Symmetrization Postulate and Its Experimental Foundation,''
  Phys.\ Rev.\  {\bf 136} (1964) B248.
  doi:10.1103/PhysRev.136.B248
  %%CITATION = doi:10.1103/PhysRev.136.B248;%%
  %81 citations counted in INSPIRE as of 10 Sep 2017

%\cite{Greenberg:1963kk}
\bibitem{Greenberg:1963kk}
  O.~W.~Greenberg and A.~M.~L.~Messiah,
  %``Selection rules for parafields and the absence of para particles in nature,''
  Phys.\ Rev.\  {\bf 138} (1965) B1155.
  doi:10.1103/PhysRev.138.B1155
  %%CITATION = doi:10.1103/PhysRev.138.B1155;%%
  %137 citations counted in INSPIRE as of 10 Sep 2017

\bibitem{Gentile}
G. Gentile, NuovoCimento {\bf 17}, 493(1940)

%\cite{Green:1952kp}
\bibitem{Green:1952kp}
  H.~S.~Green,
  %``A Generalized method of field quantization,''
  Phys.\ Rev.\  {\bf 90} (1953) 270.
  doi:10.1103/PhysRev.90.270
  %%CITATION = doi:10.1103/PhysRev.90.270;%%
  %438 citations counted in INSPIRE as of 10 Sep 2017

%\cite{Ignatiev:1987zd}
\bibitem{Ignatiev:1987zd}
  A.~Y.~Ignatiev and V.~A.~Kuzmin,
  %``Search for Small Violation of the Pauli Principle,''
  JETP Lett.\  {\bf 47} (1988) 4.
  %%CITATION = JTPLA,47,4;%%
  %9 citations counted in INSPIRE as of 10 Sep 2017
  
  
 %\cite{Gavrin:1988nu}
\bibitem{Gavrin:1988nu}
  V.~N.~Gavrin, A.~Y.~Ignatiev and V.~A.~Kuzmin,
  %``Search for Small Violation of the Pauli Principle,''
  Phys.\ Lett.\ B {\bf 206} (1988) 343.
  doi:10.1016/0370-2693(88)91518-3
  %%CITATION = doi:10.1016/0370-2693(88)91518-3;%%
  %16 citations counted in INSPIRE as of 10 Sep 2017


\bibitem{PEP1}
O.W. Greenberg, Phys. Rev. Lett. {\bf 64}, 705 (1990)
\bibitem{PEP2}
R.N. Mohapatra, Phys. Lett. B {\bf 242}, 407 (1990)

\bibitem{PEP3}
O.W.Greenberg, R.C.Hilborn, Fund.Phys.{\bf 29}, 397(1999)


  %\cite{Collins:2004bp}
\bibitem{Collins:2004bp} 
  J.~Collins, A.~Perez, D.~Sudarsky, L.~Urrutia and H.~Vucetich,
  %``Lorentz invariance and quantum gravity: an additional fine-tuning problem?,''
  Phys.\ Rev.\ Lett.\  {\bf 93}, 191301 (2004)
  doi:10.1103/PhysRevLett.93.191301
  [gr-qc/0403053].
  %%CITATION = doi:10.1103/PhysRevLett.93.191301;%%
  %232 citations counted in INSPIRE as of 15 Sep 2017
  

%NONCOMMUTATIVE:

\bibitem{Heisenberg}
R. Jackiw, Nucl.Phys.Proc.Suppl. 108, 30 (2002) [hep-th/0110057]; Letter of Heisenberg to Peierls (1930), Wolfgang Pauli, Scientific Correspondence, Vol. II, p.15, Ed. Karl von Meyenn, Springer-Verlag, 1985; Letter of Pauli to Oppenheimer (1946), Wolfgang Pauli, Scientific Correspondence, Vol. III, p.380, Ed. Karl von Meyenn, Springer-Verlag, 1993.

\bibitem{Snyder}
H. Snyder, Phys. Rev. {\bf 71}, 38 (1947).

\bibitem{Yang}
C. N. Yang, Phys. Rev. {\bf 72}, 874 (1947).

  %%LOOPS & STRINGS
  
  %\cite{Addazi:2017qwt}
\bibitem{Addazi:2017qwt}
  A.~Addazi and A.~Marciano,
  %``A new duality between Topological M-theory and Loop Quantum Gravity,''
  arXiv:1707.05347 [hep-th].
  %%CITATION = ARXIV:1707.05347;%%
  

%NON-COMMUTATIVE STRING THEORY 

%\cite{Frohlich:1993es}
\bibitem{Frohlich:1993es}
  J.~Frohlich and K.~Gawedzki,
  %``Conformal field theory and geometry of strings,''
  In *Vancouver 1993, Proceedings, Mathematical quantum theory, vol. 1* 57-97, and Preprint - Gawedzki, K. (rec.Nov.93) 44 p
  [hep-th/9310187].
  %%CITATION = HEP-TH/9310187;%%
  %67 citations counted in INSPIRE as of 11 Sep 2017
  
  \bibitem{Chamseddine1}
  A.H. Chamseddine and J. Fr\"ohlich, {\it Some Elements of Connes? Noncommutative Geometry and Space-time Geometry},
   in: {\it Yang Festschrift}, eds. C.S. Liu and S.-F. Yau (International Press, Boston, 1995) 10-34.
   
   %\cite{Frohlich:1995mr}
\bibitem{Frohlich:1995mr}
  J.~Frohlich, O.~Grandjean and A.~Recknagel,
  %``Supersymmetric quantum theory, noncommutative geometry, and gravitation,''
  In *Les Houches 1995, Quantum symmetries* 221-385
  [hep-th/9706132].
  %%CITATION = HEP-TH/9706132;%%
  %51 citations counted in INSPIRE as of 11 Sep 2017
  
  %\cite{Connes:1997cr}
\bibitem{Connes:1997cr}
  A.~Connes, M.~R.~Douglas and A.~S.~Schwarz,
  %``Noncommutative geometry and matrix theory: Compactification on tori,''
  JHEP {\bf 9802} (1998) 003
  doi:10.1088/1126-6708/1998/02/003
  [hep-th/9711162].
  %%CITATION = doi:10.1088/1126-6708/1998/02/003;%%
  %1635 citations counted in INSPIRE as of 11 Sep 2017
  
  %\cite{Seiberg:1999vs}
\bibitem{Seiberg:1999vs}
  N.~Seiberg and E.~Witten,
  %``String theory and noncommutative geometry,''
  JHEP {\bf 9909} (1999) 032
  doi:10.1088/1126-6708/1999/09/032
  [hep-th/9908142].
  %%CITATION = doi:10.1088/1126-6708/1999/09/032;%%
  %3819 citations counted in INSPIRE as of 11 Sep 2017
   
  
  %NON-COMMUTATIVE LQG

%\cite{AmelinoCamelia:2003xp}
\bibitem{AmelinoCamelia:2003xp}
  G.~Amelino-Camelia, L.~Smolin and A.~Starodubtsev,
  %``Quantum symmetry, the cosmological constant and Planck scale phenomenology,''
  Class.\ Quant.\ Grav.\  {\bf 21} (2004) 3095
  doi:10.1088/0264-9381/21/13/002
  [hep-th/0306134].
  %%CITATION = doi:10.1088/0264-9381/21/13/002;%%
  %182 citations counted in INSPIRE as of 11 Sep 2017



%\cite{Freidel:2005me}
\bibitem{Freidel:2005me}
  L.~Freidel and E.~R.~Livine,
  %``Effective 3-D quantum gravity and non-commutative quantum field theory,''
  Phys.\ Rev.\ Lett.\  {\bf 96} (2006) 221301
  doi:10.1103/PhysRevLett.96.221301
  [hep-th/0512113].
  %%CITATION = doi:10.1103/PhysRevLett.96.221301;%%
  %202 citations counted in INSPIRE as of 11 Sep 2017
  
  %\cite{Cianfrani:2016ogm}
\bibitem{Cianfrani:2016ogm}
  F.~Cianfrani, J.~Kowalski-Glikman, D.~Pranzetti and G.~Rosati,
  %``Symmetries of quantum spacetime in three dimensions,''
  Phys.\ Rev.\ D {\bf 94} (2016) no.8,  084044
  doi:10.1103/PhysRevD.94.084044
  [arXiv:1606.03085 [hep-th]].
  %%CITATION = doi:10.1103/PhysRevD.94.084044;%%
  %13 citations counted in INSPIRE as of 11 Sep 2017

%\cite{Amelino-Camelia:2016gfx}
\bibitem{Amelino-Camelia:2016gfx}
  G.~Amelino-Camelia, M.~M.~da Silva, M.~Ronco, L.~Cesarini and O.~M.~Lecian,
  %``Spacetime-noncommutativity regime of Loop Quantum Gravity,''
  Phys.\ Rev.\ D {\bf 95} (2017) no.2,  024028
  doi:10.1103/PhysRevD.95.024028
  [arXiv:1605.00497 [gr-qc]].
  %%CITATION = doi:10.1103/PhysRevD.95.024028;%%
  %15 citations counted in INSPIRE as of 11 Sep 2017
  
\bibitem{NClimLQG}
G. Amelino-Camelia, M. M. da Silva, M. Ronco, L. Cesarini, O. M. Lecian, 
``Spacetime-noncommutativity regime of Loop Quantum Gravity,''
Phys. Rev. D {\bf 95}, 024028 (2017) [arXiv:1605.00497].

\bibitem{BRAM}
S. Brahma, M. Ronco, G. Amelino-Camelia and A. Marciano, ``Linking loop quantum gravity quantization ambiguities with phenomenology,'' 
Phys. Rev. D {\bf 95}, 4 (044005)2017 [arXiv:1610.07865].
  
  %\cite{Brahma:2017yza}
\bibitem{Brahma:2017yza} 
  S.~Brahma, A.~Marciano and M.~Ronco,
  %``Quantum position operator: why space-time lattice is fuzzy,''
  arXiv:1707.05341 [hep-th].
  %%CITATION = ARXIV:1707.05341;%%
  %1 citations counted in INSPIRE as of 15 Sep 2017


\bibitem{MarMod}
S. Alexander, A. Marciano and L. Modesto, 
``The Hidden Quantum Groups Symmetry of Super-renormalizable Gravity,'' Phys. Rev. D {\bf 85}, 124030 (2012) [arXiv:1202.1824 [hep-th]].  


  \bibitem{Arzano:2007ef}
M. Arzano and A. Marciano, ``Fock space, quantum fields and $\kappa$-Poincar\'e symmetries,'' Phys. Rev. D {\bf 76}, 125005 (2007) [arXiv:0707.1329].


\bibitem{AmelinoCamelia:2007uy}
G. Amelino-Camelia, G. Gubitosi, A. Marciano, P. Martinetti and F. Mercati, ``A No-pure-boost uncertainty principle from spacetime noncommutativity,'' Phys. Lett. B {\bf 671}, 
298 (2009) [arXiv:0707.1863].

\bibitem{AmelinoCamelia:2007vj}
G. Amelino-Camelia, A. Marciano and D. Pranzetti, ``On the 5D differential calculus and translation transformations in 4D kappa-Minkowski 
noncommutative spacetime,'' Int. J. Mod. Phys. A {\bf 24}, 5445 (2009) [arXiv:0709.2063].

\bibitem{AmelinoCamelia:2007wk}
G. Amelino-Camelia, F. Briscese, G. Gubitosi, A. Marciano, P. Martinetti and F. Mercati, ``Noether analysis of the twisted Hopf symmetries of 
canonical noncommutative spacetimes,'' Phys. Rev. D {\bf 78}, 025005 (2008) [arXiv:0709.4600].

\bibitem{AmelinoCamelia:2007rn}
G. Amelino-Camelia, G. Gubitosi, A. Marciano, P. Martinetti, F. Mercati, D. Pranzetti and R. A. Tacchi, ``First results of the Noether theorem 
for Hopf-algebra spacetime symmetries,'' Prog. Theor. Phys. Suppl. {\bf 171} 65, 2007 [arXiv:0710.1219].

\bibitem{Marciano:2008tva}
A. Marciano, ``On the emergence of non locality for quantum fields enjoying $\kappa$-Poincar\'e symmetries,'' Arabian J. Sci. Eng. {\bf 33}, 365 (2008).

\bibitem{Marciano:2010jm} 
A. Marciano, ``A brief overview of quantum field theory with deformed symmetries and their relation with quantum gravity,'' [arXiv:1003.0395].

\bibitem{Agostini:2006nc}
A. Agostini, G. Amelino-Camelia, M. Arzano, A. Marciano and R. A. Tacchi, ``Generalizing the Noether theorem for Hopf-algebra spacetime symmetries,'' 
Mod. Phys. Lett. A {\bf 22}, 1779 (2007) [arXiv:hep-th/0607221].

\bibitem{Arzano:2007gr}
M. Arzano and A. Marciano, ``Symplectic geometry and Noether charges for Hopf algebra space-time symmetries,'' Phys. Rev. D {\bf 75}, 
081701 (2007) [arXiv:hep-th/0701268].


\bibitem{AmelinoCamelia:2007zzb}
G. Amelino-Camelia, M. Arzano and A. Marciano, ``On the quantum-gravity phenomenology of multiparticle states,'' Frascati Phys. Ser. {\bf 43}, 155 (2007).

\bibitem{Marciano:2010gq}
A. Marciano, G. Amelino-Camelia, N. R. Bruno, G. Gubitosi, G. Mandanici and A. Melchiorri, ``Interplay between curvature and Planck-scale effects 
in astrophysics and cosmology,'' JCAP {\bf 1006} 030, 2010 [arXiv:1004.1110].

\bibitem{AmelinoCamelia:2010zf}
G. Amelino-Camelia, A. Marciano, M. Matassa and G. Rosati, ``Testing quantum-spacetime relativity with gamma-ray telescopes'' [arXiv:1006.0007].

\bibitem{AmelinoCamelia:2012it}
G. Amelino-Camelia, A. Marciano, M. Matassa and G. Rosati, ``Deformed Lorentz symmetry and relative locality in a curved/expanding spacetime,'' 
Phys. Rev. D {\bf 86}, 124035 (2012) [arXiv:1206.5315].



 %\cite{AlvarezGaume:2003mb}
\bibitem{AlvarezGaume:2003mb} 
  L.~Alvarez-Gaume and M.~A.~Vazquez-Mozo,
  %``General properties of noncommutative field theories,''
  Nucl.\ Phys.\ B {\bf 668}, 293 (2003)
  doi:10.1016/S0550-3213(03)00582-0
  [hep-th/0305093].
  %%CITATION = doi:10.1016/S0550-3213(03)00582-0;%%
  %153 citations counted in INSPIRE as of 15 Sep 2017



%QUANTUM GROUPS




%\cite{Majid:1996kd}
\bibitem{Majid:1996kd}
  S.~Majid,
  ``Foundations of quantum group theory,''
  (CUP 1995, Cambridge).
  %%CITATION = INSPIRE-427381;%%
  %17 citations counted in INSPIRE as of 10 Sep 2017

%\cite{Oeckl:2000eg}
\bibitem{Oeckl:2000eg}
  R.~Oeckl,
  %``Untwisting noncommutative R**d and the equivalence of quantum field theories,''
  Nucl.\ Phys.\ B {\bf 581} (2000) 559
  doi:10.1016/S0550-3213(00)00281-9
  [hep-th/0003018].
  %%CITATION = doi:10.1016/S0550-3213(00)00281-9;%%
  %160 citations counted in INSPIRE as of 10 Sep 2017

%\cite{Chaichian:2004za}
\bibitem{Chaichian:2004za}
  M.~Chaichian, P.~P.~Kulish, K.~Nishijima and A.~Tureanu,
  %``On a Lorentz-invariant interpretation of noncommutative space-time and its implications on noncommutative QFT,''
  Phys.\ Lett.\ B {\bf 604} (2004) 98
  doi:10.1016/j.physletb.2004.10.045
  [hep-th/0408069].
  %%CITATION = doi:10.1016/j.physletb.2004.10.045;%%
  %445 citations counted in INSPIRE as of 10 Sep 2017
  
  %\cite{Aschieri:2005yw}
\bibitem{Aschieri:2005yw}
  P.~Aschieri, C.~Blohmann, M.~Dimitrijevic, F.~Meyer, P.~Schupp and J.~Wess,
  %``A Gravity theory on noncommutative spaces,''
  Class.\ Quant.\ Grav.\  {\bf 22} (2005) 3511
  doi:10.1088/0264-9381/22/17/011
  [hep-th/0504183].
  %%CITATION = doi:10.1088/0264-9381/22/17/011;%%
  %349 citations counted in INSPIRE as of 10 Sep 2017
  
  %UNITARITY
  
  %\cite{Balachandran:2004rq}
\bibitem{Balachandran:2004rq}
  A.~P.~Balachandran, T.~R.~Govindarajan, C.~Molina and P.~Teotonio-Sobrinho,
  %``Unitary quantum physics with time-space noncommutativity,''
  JHEP {\bf 0410} (2004) 072
  doi:10.1088/1126-6708/2004/10/072
  [hep-th/0406125].
  %%CITATION = doi:10.1088/1126-6708/2004/10/072;%%
  %62 citations counted in INSPIRE as of 11 Sep 2017


 %\cite{Arzano:2016egk}
\bibitem{Arzano:2016egk} 
  M.~Arzano and J.~Kowalski-Glikman,
  %``Deformed discrete symmetries,''
  Phys.\ Lett.\ B {\bf 760}, 69 (2016)
  doi:10.1016/j.physletb.2016.06.048
  [arXiv:1605.01181 [hep-th]].
  %%CITATION = doi:10.1016/j.physletb.2016.06.048;%%
  %1 citations counted in INSPIRE as of 01 Oct 2017 
  

%SPIN VIOLATION AND MOYAL PROD

%\cite{Balachandran:2005eb}
\bibitem{Balachandran:2005eb}
  A.~P.~Balachandran, G.~Mangano, A.~Pinzul and S.~Vaidya,
  %``Spin and statistics on the Groenwald-Moyal plane: Pauli-forbidden levels and transitions,''
  Int.\ J.\ Mod.\ Phys.\ A {\bf 21} (2006) 3111
  doi:10.1142/S0217751X06031764
  [hep-th/0508002].
  %%CITATION = doi:10.1142/S0217751X06031764;%%
  %139 citations counted in INSPIRE as of 10 Sep 2017

  %\cite{Chaichian:2011fc}
\bibitem{Chaichian:2011fc} 
  M.~Chaichian, A.~D.~Dolgov, V.~A.~Novikov and A.~Tureanu,
  %``CPT Violation Does Not Lead to Violation of Lorentz Invariance and Vice Versa,''
  Phys.\ Lett.\ B {\bf 699}, 177 (2011)
  doi:10.1016/j.physletb.2011.03.026
  [arXiv:1103.0168 [hep-th]].
  %%CITATION = doi:10.1016/j.physletb.2011.03.026;%%
  %46 citations counted in INSPIRE as of 01 Oct 2017
  
  %\cite{Chaichian:2002vw}
\bibitem{Chaichian:2002vw} 
  M.~Chaichian, K.~Nishijima and A.~Tureanu,
  %``Spin statistics and CPT theorems in noncommutative field theory,''
  Phys.\ Lett.\ B {\bf 568}, 146 (2003)
  doi:10.1016/j.physletb.2003.06.009
  [hep-th/0209008].
  %%CITATION = doi:10.1016/j.physletb.2003.06.009;%%
  %95 citations counted in INSPIRE as of 01 Oct 2017
  



%%%%%%%%%%%%%%%%%%%%%%%%%%% EXPERIMENTAL limits:   %%%%%%%%%%%%%%%%%%%%%%%%%%%%%%%%%%%%%%%%%%%%%%%%%%%%%%%
%

% VIP2011
%\cite{Pichler:2016xqc}
\bibitem{Pichler:2016xqc}
  A.~Pichler {\it et al.},
  %``Application of photon detectors in the VIP2 experiment to test the Pauli Exclusion Principle,''
  J.\ Phys.\ Conf.\ Ser.\  {\bf 718} (2016) no.5,  052030
  doi:10.1088/1742-6596/718/5/052030
  [arXiv:1602.00898 [physics.ins-det]].
  %%CITATION = doi:10.1088/1742-6596/718/5/052030;%%
  %3 citations counted in INSPIRE as of 10 Sep 2017

% Ejiri 1992
\bibitem{Ejiri:1992} H. Ejiri {\it et al.}, Nucl.\ Phys.\ B (Proc.\ Suppl.) {\bf 28A} (1992) 219.

% DAMA/LIBRA 2009
%\cite{Bernabei:2009zzb}
\bibitem{Bernabei:2009zzb}
  R.~Bernabei {\it et al.},
  %``New search for processes violating the Pauli exclusion principle in sodium and in iodine,''
  Eur.\ Phys.\ J.\ C {\bf 62} (2009) 327.
  doi:10.1140/epjc/s10052-009-1068-1
  %%CITATION = doi:10.1140/epjc/s10052-009-1068-1;%%
  %82 citations counted in INSPIRE as of 10 Sep 2017

% MALBEK 2016
%\cite{Abgrall:2016wtk}
\bibitem{Abgrall:2016wtk}
  N.~Abgrall {\it et al.},
  %``Search for Pauli exclusion principle violating atomic transitions and electron decay with a p-type point contact germanium detector,''
  Eur.\ Phys.\ J.\ C {\bf 76} (2016) no.11,  619
  doi:10.1140/epjc/s10052-016-4467-0
  [arXiv:1610.06141 [nucl-ex]].
  %%CITATION = doi:10.1140/epjc/s10052-016-4467-0;%%
  %3 citations counted in INSPIRE as of 10 Sep 2017

% BOREXINO 2011
\bibitem{Bellini:2010} G. Bellini {\it et al.} (Borexino Collaboration), Phys.\ Rev.\ C {\bf 81} (2010) 034317.

% old BOREXINO
%\cite{Back:2004hd}
%\bibitem{Back:2004hd}
%  H.~O.~Back {\it et al.} [Borexino Collaboration],
%  %``New experimental limits on violations of the Pauli exclusion principle obtained with the Borexino counting test facility,''
%  Eur.\ Phys.\ J.\ C {\bf 37} (2004) 421
%  doi:10.1140/epjc/s2004-01991-1
%  [hep-ph/0406252].
%  %%CITATION = doi:10.1140/epjc/s2004-01991-1;%%
%  %54 citations counted in INSPIRE as of 10 Sep 2017


% KAMIOKANDE 1992
%\cite{Suzuki:1993zp}
\bibitem{Suzuki:1993zp}
  Y.~Suzuki {\it et al.} [Kamiokande Collaboration],
  %``Study of invisible nucleon decay, N ---> neutrino neutrino anti-neutrino, and a forbidden nuclear transition in the Kamiokande detector,''
  Phys.\ Lett.\ B {\bf 311} (1993) 357.
  doi:10.1016/0370-2693(93)90582-3
  %%CITATION = doi:10.1016/0370-2693(93)90582-3;%%
  %37 citations counted in INSPIRE as of 10 Sep 2017

\bibitem{bernabei2} R.~Bernabei {\it et al.}, Phys.\ Lett.\ B {\bf 460} (1999) 236.

%
%%%%%%%%%%%%%%%%%%%%%%%%%%%%%%%%%%%%%%%%%%%%%%%%%%%%%%%%%%%%%%%%%%%%%%%%%%%%%%%%%%%%%%%%%%%%%%%%%%%%%%%%%%


  
%\cite{Gayral:2003dm}
\bibitem{Gayral:2003dm} 
  V.~Gayral, J.~M.~Gracia-Bondia, B.~Iochum, T.~Schucker and J.~C.~Varilly,
  %``Moyal planes are spectral triples,''
  Commun.\ Math.\ Phys.\  {\bf 246}, 569 (2004)
  doi:10.1007/s00220-004-1057-z
  [hep-th/0307241].
  %%CITATION = doi:10.1007/s00220-004-1057-z;%%
  %105 citations counted in INSPIRE as of 05 Oct 2017
  
  %\cite{Iochum:2010hi}
\bibitem{Iochum:2010hi} 
  B.~Iochum, T.~Masson, T.~Schucker and A.~Sitarz,
  %``Compact $\kappa$-Deformation and Spectral Triples,''
  Rept.\ Math.\ Phys.\  {\bf 68}, 37 (2011)
  doi:10.1016/S0034-4877(11)60026-8
  [arXiv:1004.4190 [hep-th]].
  %%CITATION = doi:10.1016/S0034-4877(11)60026-8;%%
  %10 citations counted in INSPIRE as of 05 Oct 2017
  
  %\cite{Matassa:2012su}
\bibitem{Matassa:2012su} 
  M.~Matassa,
  %``A modular spectral triple for $\kappa$-Minkowski space,''
  J.\ Geom.\ Phys.\  {\bf 76}, 136 (2014)
  doi:10.1016/j.geomphys.2013.10.023
  [arXiv:1212.3462 [math-ph]].
  %%CITATION = doi:10.1016/j.geomphys.2013.10.023;%%
  %3 citations counted in INSPIRE as of 05 Oct 2017
  

  
  %\cite{Arzano:2007nx}
\bibitem{Arzano:2007nx} 
  M.~Arzano,
  %``Quantum fields, non-locality and quantum group symmetries,''
  Phys.\ Rev.\ D {\bf 77}, 025013 (2008)
  doi:10.1103/PhysRevD.77.025013
  [arXiv:0710.1083 [hep-th]].
  %%CITATION = doi:10.1103/PhysRevD.77.025013;%%
  %27 citations counted in INSPIRE as of 04 Oct 2017

%\cite{Arzano:2008yc}
\bibitem{Arzano:2008yc} 
  M.~Arzano, A.~Hamma and S.~Severini,
  %``Hidden entanglement at the Planck scale: Loss of unitarity and the information paradox,''
  Mod.\ Phys.\ Lett.\ A {\bf 25}, 437 (2010)
  doi:10.1142/S0217732310032603
  [arXiv:0806.2145 [hep-th]].
  %%CITATION = doi:10.1142/S0217732310032603;%%
  %8 citations counted in INSPIRE as of 25 Sep 2017
  
  %\cite{Freidel:2007hk}
\bibitem{Freidel:2007hk} 
  L.~Freidel, J.~Kowalski-Glikman and S.~Nowak,
  %``Field theory on kappa-Minkowski space revisited: Noether charges and breaking of Lorentz symmetry,''
  Int.\ J.\ Mod.\ Phys.\ A {\bf 23}, 2687 (2008)
  doi:10.1142/S0217751X08040421
  [arXiv:0706.3658 [hep-th]].
  %%CITATION = doi:10.1142/S0217751X08040421;%%
  %60 citations counted in INSPIRE as of 25 Sep 2017

%\cite{Daszkiewicz:2007ru}
\bibitem{Daszkiewicz:2007ru} 
  M.~Daszkiewicz, J.~Lukierski and M.~Woronowicz,
  %``Towards quantum noncommutative kappa-deformed field theory,''
  Phys.\ Rev.\ D {\bf 77}, 105007 (2008)
  doi:10.1103/PhysRevD.77.105007
  [arXiv:0708.1561 [hep-th]].
  %%CITATION = doi:10.1103/PhysRevD.77.105007;%%
  %67 citations counted in INSPIRE as of 25 Sep 2017


  
    %\cite{ArkaniHamed:1998nn}
  \bibitem{ArkaniHamed:1998nn} 
  N.~Arkani-Hamed, S.~Dimopoulos and G.~R.~Dvali,
  %``Phenomenology, astrophysics and cosmology of theories with submillimeter dimensions and TeV scale quantum gravity,''
  Phys.\ Rev.\ D {\bf 59}, 086004 (1999)
  doi:10.1103/PhysRevD.59.086004
  [hep-ph/9807344].
  %%CITATION = doi:10.1103/PhysRevD.59.086004;%%
  %2560 citations counted in INSPIRE as of 04 Oct 2017
  
  %\cite{ArkaniHamed:1998rs}
\bibitem{ArkaniHamed:1998rs} 
  N.~Arkani-Hamed, S.~Dimopoulos and G.~R.~Dvali,
  %``The Hierarchy problem and new dimensions at a millimeter,''
  Phys.\ Lett.\ B {\bf 429}, 263 (1998)
  doi:10.1016/S0370-2693(98)00466-3
  [hep-ph/9803315].
  %%CITATION = doi:10.1016/S0370-2693(98)00466-3;%%
  %6099 citations counted in INSPIRE as of 04 Oct 2017
  
  %\cite{Dvali:2001gm}
\bibitem{Dvali:2001gm} 
  G.~R.~Dvali, G.~Gabadadze, M.~Kolanovic and F.~Nitti,
  %``The Power of brane induced gravity,''
  Phys.\ Rev.\ D {\bf 64}, 084004 (2001)
  doi:10.1103/PhysRevD.64.084004
  [hep-ph/0102216].
  %%CITATION = doi:10.1103/PhysRevD.64.084004;%%
  %179 citations counted in INSPIRE as of 04 Oct 2017
  
%\bibitem{juno}  V. Antonelli and L. Miramonti, arXiv:1710.07401.

\bibitem{C1}
E.~Costa et al., %``Discovery of the X-ray afterglow of the ??-ray burst of 28 February 1997Ó,
Nature, 387, 783-785 (1997) [arXiv:astro-ph/9706065 [astro-ph]]. 

%\bibitem{C}
\bibitem{C2}
G.~Amelino-Camelia, J.R.~Ellis,  N.E.~Mavromatos, D.V.~Nanopoulos and S.~Sarkar, %``Potential Sensitivity of Gamma-Ray Burster Observations to Wave Dispersion in VacuoÓ, 
Nature, 393, 763-765 (1998) [arXiv:astro-ph/9712103].

\bibitem{C3}
J.~Granot (Fermi LAT and GBM collaborations), %``GRB Theory in the Fermi EraÓ, 
p.321-328 of the Proceedings of the {\it 44th Rencontres de Moriond on Very High Energy Phenomena in the Universe}, eConf C09-02-01.2 [arXiv:0905.2206 [astro-ph.HE]].

\bibitem{C4}
G.~Amelino-Camelia and L.~Smolin, %``Prospects for constraining quantum gravity dispersion with near term observationsÓ, 
Phys. Rev. {\bf D} 80, 084017 (2009) [arXiv:0906.3731 [astro-ph.HE]].

\bibitem{C5}
R.J.~Nemiro, J.~Holmes, and R.~Connolly, %``Bounds on Spectral Dispersion from Fermi-detected Gamma Ray BurstsÓ, 
Phys. Rev. Lett., 108, 231103 (2012) [arXiv:1109.5191 [astro-ph.CO]].
  
\bibitem{C6}
G. Amelino-Camelia, %``Quantum-Spacetime PhenomenologyÓ,
Living Rev. Rel. 16 (2013) 5  [arXiv:0806.0339].   

\bibitem{C7}  
J.~Albert et al. (MAGIC Collaboration), %``Probing quantum gravity using photons from a flare of the active galactic nucleus Markarian 501 observed by the MAGIC telescopeÓ, 
Phys. Lett. {\bf B}, 668, 253-257 (2008) [arXiv:0708.2889 [astro-ph]].

\bibitem{C8}  
F.~Aharonian et al. (HESS Collaboration), %``Limits on an Energy Dependence of the Speed of Light from a Flare of the Active Galaxy PKS 2155-304Ó, 
Phys. Rev. Lett., 101, 170402 (2008) [arXiv:0810.3475 [astro-ph]].
  
\bibitem{C9}   
G.~Ghirlanda, G.~Ghisellini and L.~Nava, %``The onset of the GeV afterglow of GRB 090510Ó, 
Astron. Astrophys. 510 (2010) L7, [arXiv:0909.0016 [astro-ph.HE]]. 

\bibitem{C10}  
A.A.~Abdo et al. (Fermi LAT and Fermi GBM), %``A limit on the variation of the speed of light arising from quantum gravity effectsÓ, 
Nature, 462, 331-334 (2009).

\bibitem{C11}  
A.~Abramowski et al. (HESS Collaboration), %``Search for Lorentz invariance breaking with a likelihood t of the PKS 2155-304 are data taken on MJD 53944Ó, 
Astropart. Phys., 34, 738-747 (2011) [arXiv:1101.3650 [astro-ph.HE]].
  
\bibitem{C12} 
O.~Bertolami and C.S.~Carvalho, %``Proposed astrophysical test of Lorentz invarianceÓ, 
Phys. Rev. {\bf D} 61, 103002 (2000) [arXiv:gr-qc/9912117].
 
\bibitem{C13}  
G.~Amelino-Camelia, %ÒProposal of a second generation of quantum-gravity-motivated Lorentz- symmetry tests: Sensitivity to effects suppressed quadratically by the Planck scaleÓ,
Int. J. Mod. Phys. {\bf D} 12, 1633Ð1640 (2003) [arXiv:gr-qc/0305057].

\bibitem{C14} 
P.~M\'esz\'aros, S.~Kobayashi, S.~Razzaque and B.~Zhang, %``High energy photons, neutrinos and gravitational waves from gamma-ray burstsÓ, 
in R.~Ouyed ed., Proceedings of the First Niels Bohr Summer Institute on Beaming and Jets in Gamma Ray Bursts (NBSI), Copenhagen, Denmark, August 12 - 30, 2002, eConf C0208122 (Stanford University, Stanford, 2003) [arXiv:astro-ph/0305066]. 
%URL (accessed 10 June 2013): http://www.slac.stanford.edu/econf/C0208122/ .

\bibitem{C15} 
E.~Waxman, %``Neutrino astronomy and gamma-ray bursts'', 
Philos. Trans. R. Soc. London, Ser. {\bf A} 365, 1323-1334 (2007) [arXiv:astro-ph/0701170]. 

\bibitem{C16} 
U.~Jacob and T.~Piran, %``Neutrinos from gamma-ray bursts as a tool to explore quantum-gravity- induced Lorentz violationÓ, 
Nature Phys., 3, 87-90 (2007).

\bibitem{C17} 
G.~Amelino-Camelia and L.~Smolin, %``Prospects for constraining quantum gravity dispersion with near term observationsÓ, 
Phys. Rev. {\bf D} 80, 084017 (2009) [arXiv:0906.3731 [astro-ph.HE]].

\bibitem{C18}
%\cite{Amelino-Camelia:2015nqa}
  G.~Amelino-Camelia, D.~Guetta and T.~Piran,
  %``Icecube Neutrinos and Lorentz Invariance Violation,''
  Astrophys.\ J.\  {\bf 806}, no. 2, 269 (2015).
  doi:10.1088/0004-637X/806/2/269
  %%CITATION = doi:10.1088/0004-637X/806/2/269;%%
  %11 citations counted in INSPIRE as of 13 Jun 2018
  
\end{thebibliography}
\end{document}